# PHYSICS IN DISCRETE SPACES (A): SPACE-TIME ORGANIZATION

Pierre Peretto
Laboratoire de Physique et Modélisation des Milieux Condensés
CNRS LPMMC Grenoble (France)
peretto.pierre@neuf.fr

Abstract:
We put forward a model of discrete physical space that can account for the structure of space-time, give an interpretation to the postulates of quantum mechanics and provide a possible explanation to the organization of the standard model of particles.

## 1. A MODEL OF SPACE-TIME

The natural phenomena are usually described in the framework of a four dimensional space. This space has three equivalent space-like components, one time-like component and it is equipped with a Minkowski metrics. Space-time has usually an ontological status that is it requires no further explanations.

However 4 the total number of dimensions, 3 the number of space-like dimensions and 1 the number of time-like dimensions are numerical experimental data. If one considers that the general purpose of physics is to build theories that account for numerical experimental data, the construction of a theory of space-time is a necessity. In this essay we put forward such a model and we explore some of its consequences.

Any physical model rests upon a number of hypotheses and one can wonder what sort of hypotheses would form the basis of a relevant theory of space-time. We do not want to make any *ad hoc* hypothesis such as in string, spin-lattice or twister theories for example. We rather build the model on three statements that we cannot reject without jeopardizing physics itself. We consider the three statements and their mathematical formalizations in turn.

### A. The universe does exist

The first statement is simply that the universe does exist that is some information can be obtained on the universe through experimental observations. Information is the key word. As a matter of fact since nothing else than information is available on the nature of the universe, at least for physicists, one can assume that information itself constitutes the fabrics of the physical world.

Information is measured in terms of an information unit or bit. A bit, here called a cosmic bit (CB), is not a signal, but the simplest physical object one can imagine. Accordingly the first hypothesis of the model writes:

Our universe as a whole is entirely made of a finite countable set of cosmic bits $a = 1, 2, \cdots, N_{CB}$. The state $\sigma_a$ of a cosmic bit is a binary variable $\sigma_a = \pm 1$ analogous to an Ising (classical) spin.

The state $\Psi$ of the universe is determined by a family of CB' states

$$\Psi = \{\sigma_a\}$$

## B. The universe is not disordered

The second statement follows from the observation that the universe is not completely disordered and, therefore, that all possible states of the universe cannot be realized. As a consequence we must assume that there exists a functional of CB's states $\Lambda(\Psi)$, called a Lagrangian, which is, at least approximately, minimized for the physically realizable states of the universe. The most general Lagrangian is written as an expansion over all possible clusters of CB's.

$$\Lambda(\Psi) = \sum_{clusters} \sum_{\langle ab...c \rangle} J_{ab\cdots c} \sigma_a \sigma_b \cdots \sigma_c$$

Nothing determines the overall orientation of $\Psi$ and therefore one must have $\Lambda(\Psi) = \Lambda(-\Psi)$. This eliminates the odd terms of $\Lambda(\Psi)$. In other ways all CB's must be treated on equal footing which compels the amplitudes of interactions of same order to be one another identical. That is for clusters implying $\omega$ CB's, $a$, $b$, ..., $c$ one has, for arbitrary $a, b,...,c$

$$|J_{ab\cdots c}| = J^{(\omega)}$$

with $\omega$ an even number. For all pairs $ab$ for example $|J_{ab}| = J^{(2)} > 0$. It is assumed that the interaction amplitude $J^{(\omega)}$ decreases very rapidly with the number $\omega$ of CB's, in particular $J^{(2)} >> J^{(4)} >> \cdots$ and we shall limit the expansion of $\Lambda(\Psi)$ to fourth order terms. The signs of interactions remain to be determined. Since no knowledge exists, $J_{ab}$ is taken as a random binary variable: $J_{ab} = \pm J^{(2)}$. Likewise $J_{abcd} = \pm J^{(4)}$. Some correlations, however, possibly exist between the signs of second order interactions $J_{ab}$ and those of fourth order interactions $J_{abcd}$. In the present essay it is assumed that the sign of $J_{abcd}$ obeys a majority rule, that is

$$\text{sign}(J_{abcd}) = -\text{sign}\left(\sum_{u,v \in (abcd)} J_{uv}\right)$$

Finally

$$\Lambda(\Psi) \cong \frac{1}{2!} \sum_{ab} \left(\pm J^{(2)}\right) \sigma_a \sigma_b + \frac{1}{4!} \sum_{abcd} \left(\pm J^{(4)}\right) \sigma_a \sigma_b \sigma_c \sigma_d$$

where the sign correlations are to be taken into account

## C. The universe is not frozen

The last statement follows from the observation that the states of the universe are never completely frozen that is order is not perfect. This implies that the CB's are subject to a degree of disorder whose amplitude is determined by a parameter $b$ called "cosmic noise". Space-time is then treated as an ordinary thermodynamic system analogous to e.g. a ferromagnetic material. It can be studied by using the tools of statistical mechanics. This is not a trivial assertion because statistical mechanics rests upon two fundamental hypotheses. The first one is the ergodic hypothesis, according to which temporal averages may be replaced by ensemble averages. Since the concept of time is not yet defined only ensemble averages may be given a physical meaning. Ergodicity is then a natural hypothesis and this makes it possible to derive the statistical properties of space from usual statistical physics techniques. In particular, according to statistical physics, the probability for space to be in state $\Psi$ is given by the following Gibbs expression

$$\rho(\Psi) = \frac{1}{Z} \exp(-b\Lambda(\Psi))$$

where $Z$ is the partition function

$$Z = \sum_{\{\Psi\}} \exp(-b\Lambda(\Psi)).$$

The second basic hypothesis of statistical mechanics is the existence of a reservoir. One may imagine that the total number of CB's is infinite and that $N_{CB}$, the number of CB's belonging to our own universe, is just a finite part of this set. Then the reservoir is made of the set of CB's not belonging to our universe.

To summarize we put forward in this essay a thermodynamic model of space-time. This model is basically discrete. It introduces three, and only three, sorts of free parameters $J^{(2)}$, $J^{(4)}$ and $b$. In the model everything of our familiar physics is, *a priori*, lost, no more space or time, no more fields, no more particles. Everything has to be rebuilt. In section II we show how space-time emerges from the model, in section III we see how it can account for the basic postulates of quantum physics and in section IV we argue that it can explain the organization of the usual (Standard) model of elementary particles.

## 2. THE ORGANIZATION OF SPACE

An introduction to this section is given in (1)

### A. World points

We strive to construct a geometry starting from the three hypotheses that form the basis of our model of discrete space. First of all it is necessary to give a meaning to the concept of point since the point is the elementary object of any geometry. Because space is assumed to be discrete a physical point, here called a world point, cannot be infinitesimally small and cannot, therefore, be identified with a mathematical point.

Let us consider a cluster $W$ of $n$ cosmic bits all each other connected through negative (ferromagnetic) binary interactions

$$J_{ab} = -J^{(2)} \quad \forall a,b \in W$$

Then, according to the majority rule, one has

$$J_{abcd} = J^{(4)} \quad \forall a,b,c,d \in W$$

A world point is a cluster that minimizes its Lagrangian

$$\Lambda(W) \cong -\frac{1}{2}J^{(2)}n^2 + \frac{1}{24}J^{(4)}n^4$$

That gives $n = (6J^{(2)}/J^{(4)})^{1/2}$. Since $J^{(4)} \ll J^{(2)}$ $n$ must be a very large number. Every cosmic bit of a world point is a close neighbour of every cosmic bit of the same point and, therefore, inside a world point all physical dimensions, space or time, are meaningless concepts. The size $l^*$ of a world point is the scale below which any metrics, the Minkowski metrics in particular, is lost. $l^*$ is also the scale where the distinction between the particles, be they fermions or bosons, disappears and therefore $l^*$ should be the scale where super symmetry theories (Susy) come into play. Accordingly the metric scale $l^*$ must be of the order of $l^* \cong 10^{-21}$ cm

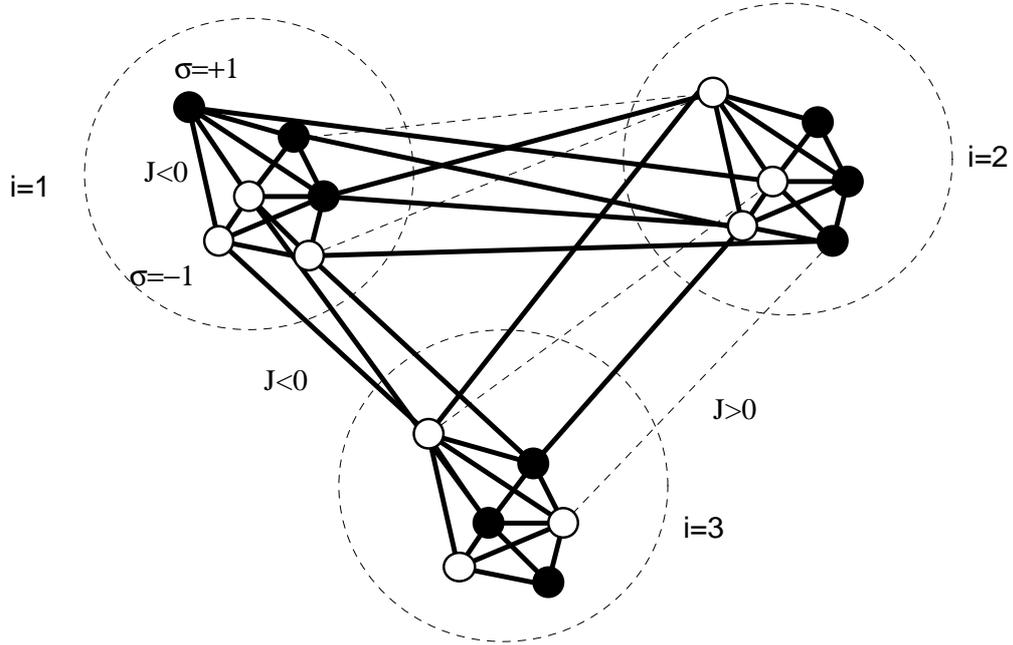

Fig 1 : The model of space we put forward in this essay. Here 18 CB's are shared between 3 world points (dotted circles) each comprised of $n = 6$ CB's. Heavy lines are for binary negative (ferromagnetic) interactions $-J^{(2)}$, dotted lines are for binary positive interactions $+J^{(2)}$.

### B. Space-time generators

Due to binary interactions the world points are polarized. The polarization $\varphi$ of a world point $W$ is defined as the thermal average $\varphi = \langle s \rangle$ of the order parameter $s$:

$$s = \frac{1}{n} \sum_{\alpha=1,\dots,n} \sigma_\alpha$$

The polarization $\varphi$ is the solution of a self consistent equation given in the mean field approach of ferromagnetism by

$$\varphi = \tanh(bJ\varphi) \qquad (1)$$

In this expression the binary interaction has been renormalized $J^{(2)} = -J/n$ so as to make the Lagrangian $\Lambda(W)$ of world point $W$ an extensive quantity. The polarization vanishes if $bJ < 1$. This situation is called symmetric vacuum. It does not vanish if $bJ > 1$ and the vacuum is asymmetric

The polarization $\varphi$ may be considered as the length of a vector $\phi$ in a $d$-dimensional abstract space called the internal space of $W$: $\phi \equiv \{\varphi_\mu\}$, $\mu = 1,\dots,d$. To give an analytical expression to the components $\varphi_\mu$ we pose the following question: can a world point be considered as a set of $d$ subsets (sub-world points so to speak) such that the system obtained by putting these $d$ sub-world points together, reproduces the polarization of the world point as a whole?

Let $n_\mu$ (with $\mu = 1,\cdots,d$), $n_\mu \cong n/d$, be the number of cosmic bits associated with a sub-world point $\mu$. The $n_\mu$'s have to obey the following constraint

$$\sum_\mu n_\mu = n$$

The order parameter of the sub-world point $\mu$ is given by

$$s_\mu = \frac{1}{n_\mu}\sum_\alpha \sigma_{\mu\alpha},$$

$\alpha = 1,\cdots,n_\mu$. The polarization of an isolated sub-world point $\mu$ is given by an equation similar to eq. (1) with a renormalized interaction

$$\langle s_\mu \rangle = \varphi_\mu = \tanh\left(b\frac{Jn_\mu}{n}\varphi_\mu\right).$$

This polarization does not vanish if

$$bJ\frac{n_\mu}{n} > 1.$$

A necessary condition for all sub-world points to be polarized at once, whatever the $n_\mu$'s, is

$$bJ\sum_\mu \frac{n_\mu}{n} = bJ > \sum_\mu 1 = d$$

The condition therefore writes $bJ > d$. It yields a highest value for $d$

$$d = \text{Int}(bJ)$$

$d$ is called the dimensionality of space. Our space is 4-dimensional. This implies that $4 < bJ < 5$, that is $bJ > 1$ and the vacuum is asymmetric indeed. The free energy of a system of $d$ sub world points is given by

$$F = \frac{-nJ}{2d^2}\sum_{\mu\nu}\varphi_\mu\varphi_\nu + \frac{n}{2bd}\sum_\mu \left((1+\varphi_\mu)\text{Ln}(1+\varphi_\mu) + (1-\varphi_\mu)\text{Ln}(1-\varphi_\mu)\right)$$

The calculation, a classical calculation in statistical mechanics, is given in Appendix A1.
The generators of space are obtained by solving the set of $d$ equations given by $\partial F/\partial \varphi_\mu = 0$
In the case where $d=1$ that is to say if

$$\varphi_\mu = \varphi\delta_{\mu 1}$$

the free energy per bit reduces to

$$F(\varphi)/n = \frac{-J}{2}\varphi^2 + \frac{1}{b}\left[\left(\frac{1+\varphi}{2}\right)\text{Ln}(1+\varphi) + \left(\frac{1-\varphi}{2}\right)\text{Ln}(1-\varphi)\right] \qquad (2)$$

The condition $\partial F/\partial \varphi = 0$ gives eq.(1). Expanding the logarithmic functions to fourth order yields a Landau type free energy

$$F(\varphi) = \lambda\varphi^2 + \mu\varphi^4$$

with

$$\lambda = \frac{(1-bJ)}{2b} \quad ; \quad \mu = \frac{1}{12b}$$

### C. World points gauge symmetries

The model involves several sorts of symmetry that can be used to give more explicit expressions to dimensions generators.

First of all one has

$$\sum_\mu \varphi_\mu = \left\langle \frac{1}{n_\mu} \sum_{\alpha\mu} \sigma_{\alpha\mu} \right\rangle = d \left\langle \frac{1}{n} \sum_{\alpha \in W} \sigma_\alpha \right\rangle = d\varphi$$

and

$$d^2\varphi^2 = \sum_\mu \varphi_\mu^2 + \sum_{\mu,\nu \neq \mu} \varphi_\mu \varphi_\nu$$

Averaging this expression over all possible rotations in the internal space of $W$ makes the cross terms vanish and the expression becomes

$$\sum_\mu \varphi_\mu^2 = d^2\varphi^2 \qquad (3)$$

which means that the polarization vector $\phi$ of $W$

$$\phi = \begin{pmatrix} \vdots \\ \varphi_\mu \\ \vdots \\ \varphi_d \end{pmatrix}$$

is normalized. This gives rise to a local gauge invariance that states that physics must be left invariant under any rotation of $\phi$ in the internal space of a world point.

In other respects the labels of the polarization components are arbitrary and, therefore, re-labelling the names of the components must not change the physical phenomena. This gives rise to another gauge invariance which states that physics must be left invariant under the operations of the group $S_d$ of permutations of $d$ objects. Locality means that both criteria of invariance must be satisfied for each world point $W$ individually.

A state $\Psi$ of the universe as a whole can be reformulated in terms of polarizations. It is written as a column vector

$$\Psi = \begin{pmatrix} \vdots \\ \phi_i \\ \vdots \\ \vdots \end{pmatrix}$$

where the state $\phi_i$ of world point $i$, $i = 1,..., N_W \ (= N_{CB}/n)$ is defined by

$$\phi_i = \begin{pmatrix} \vdots \\ \varphi_{i\mu} \\ \vdots \\ \varphi_{id} \end{pmatrix}$$

The Lagrangian (limited to second terms) is given by

$$\Lambda(\Psi) = \frac{1}{2} \sum_{i\mu\alpha, j\nu\beta} J_{i\mu\alpha, j\nu\beta} \sigma_{i\mu\alpha} \sigma_{j\nu\beta}$$

In mean field theories there are no polarization fluctuations. The free energy can then be expressed as

$$F = \langle \Lambda \rangle = \frac{1}{2} \sum_{i\mu\alpha, j\nu\beta} J_{i\mu\alpha, j\nu\beta} \langle \sigma_{i\mu\alpha} \rangle \langle \sigma_{j\nu\beta} \rangle = \sum_{i\mu, j\nu} \left( \sum_{\alpha\beta} \frac{J_{i\mu\alpha, j\nu\beta}}{2} \right) \langle s_{i\mu} \rangle \langle s_{j\mu} \rangle = \sum_{i\mu, j\nu} K_{i\mu, j\nu} \varphi_{i\mu} \varphi_{j\nu}$$

Since $F$ must be left unchanged under the permutations or rotations of polarization components the parameters $K_{i\mu, j\nu}$ factorize

$$K_{i\mu, j\nu} = \Delta_{ij} G_{\mu\nu}$$

that is

$$F = \Psi^T K \Psi$$

with

$$K = \Delta \otimes G$$

$G$ is a square, real, symmetric, $d$-dimensional matrix that operates in the internal space on the polarization components of a given cosmic point. The elements $G_{\mu\nu}$ describe the interaction between the $\mu$ and $\nu$ components of the polarization inside one and the same world point $W$.
$\Delta$ is a square, real, symmetric, matrix. An element $\Delta_{ij}$ describes the interaction that links a world point "$i$" to world point "$j$". $\Delta_{ij}$ is a sum of binary random variables $\pm J/n$. Its off-diagonal coefficients are random variables of the order of $\Delta_{ij} \cong 1/n^{1/2}$ whereas the diagonal terms are of the order of $\Delta_{ii} \cong 1$.
Permutation gauge invariance imposes $G$ to be left unchanged under any operation of the permutation group $S_d$ of $d$ objects. $G$ may be written accordingly as

$$G_{\mu\nu} = J_0 \delta_{\mu\nu} + J_1 (1 - \delta_{\mu\nu})$$

Then, the free energy of a given world point $W$ reads

$$F(W) = J_0 \sum_{\mu} \varphi_{\mu}^2 + J_1 \sum_{\mu\nu(\neq\mu)} \varphi_{\mu} \varphi_{\nu} \qquad (4)$$

To express the parameters $J_0$ and $J_1$ in terms of $b$, $J$ and $n$, the three parameters of the model, it is necessary to identify the two expressions (2) and (4) of $F$. By expanding the logarithmic functions to second order and by using the definition of polarization components, one has

$$F(W) = \left[ \frac{n}{2d^2} \sum_{\mu} (-J + d/b) \varphi_{\mu}^2 - \frac{Jn}{2d^2} \sum_{\mu\nu(\neq\mu)} \varphi_{\mu} \varphi_{\nu} \right] \qquad (5)$$

The identification of eq. (4) with eq. (5) yields

$$J_0 \cong \frac{n}{2d^2} \left( -J + \frac{d}{b} \right),$$

$$J_1 \cong -\frac{n}{2d^2} J.$$

If a convenient form of $G$ is (4), this form is not unique because any unitary transformation of this representation is also convenient, in particular the one that makes $G$ diagonal. Since

$$\text{Det}(G - \lambda I) = (J_0 + (d-1)J_1 - \lambda)(J_0 - J_1 - \lambda)^{d-1}$$

the diagonal representation identifies two and only two subspaces for $G$. The first one corresponds to the eigenvalue

$$G_{tt} = J_0 + (d-1)J_1 = \frac{n}{2db}(1 - bJ)$$

It is not degenerate. This subspace, of dimension 1 whatever *d,* will be called "time dimension". The other subspace corresponds to the eigenvalue

$$G_{rr} = J_0 - J_1 = \frac{n}{2bd}.$$

This subspace, of dimension *d*-1, corresponds to "space dimensions". In this form $G$ is called the metric matrix $G_0$. Let us write

$$G_{0\mu\mu} = \text{Sign}(G_{\mu\mu})|G_{\mu\mu}| = \varepsilon_\mu G_{0\mu}$$

and define the metric tensor *g* by

$$g_{\mu\nu} = \delta_{\mu\nu}\varepsilon_\mu$$

Since *bJ>1* its elements are

$$\varepsilon_t = \text{Sign}\left(\frac{n}{2db}(1 - Jb)\right) = -1$$

$$\varepsilon_r = \text{Sign}\left(\frac{n}{2bd}\right) = +1$$

that is

$$g = \begin{pmatrix} -1 & & & \\ & +1 & & \\ & & \ddots & \\ & & & +1 \end{pmatrix}$$

The metrics is therefore Minkowskian. It would be Euclidian if $bJ < 1$. It is worth pointing out that there is no more ambiguity on the sign of *g* (whereas relativistic mechanics does not distinguish between *g* and *-g*). The three dimensions of space and the unique dimension of time constitute a conformal space with dilatations factors given by $G_{0r}$ and $G_{0t}$ respectively.

From $G_{0r}$ and $G_{0t}$ it is possible to define two constants, c and $\hbar$, whose physical meaning will be discussed below. They are given by

$$c^2 = \left|\frac{G_{0r}}{G_{0t}}\right| = \frac{1}{bJ - 1}$$

and

$$\hbar = l * G_{0r}^{1/2} = l * (n/2bd)^{1/2}$$

The usual meaning of space and time will be recovered below when we derive the Klein-Gordon equation.

As a matter of fact, this organization of space is fully determined by the irreducible representations of groups of permutations of *d* objects. For example the permutation group $S_4$ of four objects has $4! = 24$ elements. Since $S_4$ has 5 classes there are 5 irreducible representations that are

$$\Gamma_1, \Gamma_1^*, \Gamma_2, \Gamma_3, \Gamma_3^*$$

with orders 1, 1, 2, 3 and 3 respectively. The table of characters of these representations is given in Table-I.

| classes | $(1):1$ | $(ab):6$ | $(ab)(cd):3$ | $(abc):8$ | $(abcd):6$ |
|---|---|---|---|---|---|
| $\Gamma_1$ | 1 | 1 | 1 | 1 | 1 |
| $\Gamma_1^*$ | 1 | $-1$ | 1 | 1 | $-1$ |
| $\Gamma_2$ | 2 | 0 | 2 | $-1$ | 0 |
| $\Gamma_3$ | 3 | 1 | $-1$ | 0 | $-1$ |
| $\Gamma_3^*$ | 3 | $-1$ | $-1$ | 0 | 1 |

Table-I: Table of characters of $S_4$

The invariance of four dimensional matrices, such as *G*, under those transformations, requires the matrix to commute with the 24 matrices of permutations. An example of a permutation matrix is

$$\begin{pmatrix} . & . & . & 1 \\ 1 & . & . & . \\ . & . & 1 & . \\ . & 1 & . & . \end{pmatrix}$$

which is a four dimensional representation of the permutation (1234)=>(2431). Let $\Gamma_4$ be this representation. Its characters are given in Table-II:

| classes | $(1):1$ | $(ab):6$ | $(ab)(cd):3$ | $(abc):8$ | $(abcd):6$ |
|---|---|---|---|---|---|
| $\Gamma_4$ | 4 | 2 | 0 | 1 | 0 |

Table-II. Table of characters of $\Gamma_4$

From these tables it is deduced that

$$\Gamma_4 = \Gamma_1 \oplus \Gamma_3$$

a sum of two irreducible representations with dimensions 1 and 3 respectively. The properties of the group $S_4$ are important because they fully determine the main aspects of the standard model of particles (see Section 4).

## 3. THE POSTULATES OF QUANTUM THEORY

The physically realizable states $\Psi$ of the universe are obtained by minimizing the Lagrangian $\Lambda = \Psi^T(\Delta \otimes G)\Psi$ under the constraint $\Psi^T\Psi = \sum_{i=1,...,N_W} \phi_i^T \phi_i = N_W d^2 \varphi^2$. This amounts to finding the minima, with respect to $\Psi^T$, of

$$\Psi^T(\Delta \otimes G)\Psi - \lambda(\Psi^T\Psi - N_W d^2 \varphi^2)$$

where $\lambda$ is a Lagrange multiplier. The solution is an eigenvalue equation :

$$(\Delta \otimes G)\Psi = \lambda \Psi \tag{6}$$

It is unlikely, however, that $\Psi$ really represents a physical state of the whole universe because it is unlikely for the coherence of $\Psi$ to be preserved everywhere in the universe. Equation (6) must therefore be valid only for a (small) part of the universe wherein $\Psi$ keeps its coherence. This part, comprized of *N* world points, is called a quantum system and $\psi$, the piece of $\Psi$ that belongs to the quantum system, is called a quantum state.

The model of discrete space imposes some properties to quantum states. These properties are usually expressed in the form of postulates which, therefore rather appear as consequences of the model itself.

## A. First postulate

The states $\psi$ of a quantum system are normalized. They are determined, we have seen, by the minimization of the Lagrangian

$$\Lambda(\psi) = \psi^T (\Delta \otimes G)\psi$$

under the conventional normalization condition

$$\psi^T \psi = 1$$

The condition is satisfied if the states obey the following eigenvalue equation

$$(\Delta \otimes G)\psi = \kappa\psi \qquad (7)$$

The random matrix $\Delta$ and the matrix $G$ are real-valued and symmetrical. Then the eigenvalues $\kappa$ and the eigenstates $\psi$ are real-valued.

If $\psi_1$ and $\psi_2$ are two eigenstates associated with the eigenvalue $\kappa$ the linear combination

$$\psi = a_1\psi_1 + a_2\psi_2$$

is also an eigenstates associated with $\kappa$. Therefore the set of eigenstates associated with $\kappa$ makes a (Hilbert) vector space. This space is real-valued but it can be made complex-valued by introducing local phase factors and by letting

$$\phi_i \to \phi_i^C = (\exp i\eta_i)\phi_i \ ; \ \phi_i^{CT*} = (\exp -i\eta_i)\phi_i^T$$

This does not change the normalization condition

$$\phi_i^{CT*}\phi_i^C = \phi_i^{T*}\phi_i = 1$$

and transforms the matrix $\Delta$ into

$$\Delta_{ij} \to \Delta_{ij}^C = \exp(i(\eta_i - \eta_j))\Delta_{ij}$$

an hermitian operator that has exactly the same properties as $\Delta$ (in particular the same real eigenvalues) because the eigenvalue equation

$$\sum_j \Delta_{ij}\phi_j = \kappa\phi_i$$

becomes

$$\sum_j [(\exp i\eta_i)(\Delta_{ij})\exp(-i\eta_j)]\exp(i\eta_j)\phi_j = \kappa \exp(i\eta_i)\phi_i$$

or

$$\sum_j \Delta_{ij}^C \phi_j^C = \kappa\phi_i^C$$

Transforming the real vector space into a complex vector space therefore has no effect whatsoever as far as physics is concerned.

The first postulate writes accordingly:
The states of a quantum system constitute a Hilbert space, that is to say a complex vector space equipped with an inner product that is defined for all its vectors. Let $\psi$ and $\chi$ be two states of a quantum system. These states can be normalized

$$\psi^T * \psi = \chi^T * \chi = 1$$

and the linear superposition

$$\varphi = \frac{\lambda\psi + \mu\chi}{|\lambda\psi + \mu\chi|}$$

is an allowed state of the quantum system.

Remark
Linearity is the most central and striking property of quantum mechanics. In the present approach, linearity stems from the quadratic form of the Lagrangian $\Lambda$ which is a result of limiting $\Lambda$ to binary interactions. As far as quantum theory is concerned the influence of fourth order interactions is completely negligible.

## B. Second postulate

Besides local symmetries that reflect the symmetry properties of world points, a quantum system may also display global symmetries.

Global symmetries are operations carried out on a quantum system as a whole that leave the physics of the system unchanged. The set of such operations constitutes a finite group P that permute the $n$ labels '$i,...,j$' of world points. We recall that, according to Cayley theorem, any finite group can be considered as a sub-group of a permutation group. One may associate a linear operator $A^P$ to a global symmetry P by compelling the operator to remain unchanged under the operations $\omega^P$ of P.

$$A^P = \omega^P A^P \omega^{P-1} \quad \forall \, \omega^P \in P$$

The linear operator $A^P$ reflects the physical (symmetry) properties of the system. It may be obtained by projecting the Lagrangian $\Lambda$ on the trivial representation of P.

$$A^P = \frac{1}{n_P} \sum_P \omega^P \Lambda \omega^{P-1}$$

where $n_P$ is the order of P. Let $\lambda$ be an eigenvalue of $\Lambda$: $\Lambda \psi_\Lambda = \lambda \psi_\Lambda$. $\lambda$ is also an eigenvalue of each term of $A^P$ since

$$(\omega^P \Lambda \omega^{P-1})(\omega^P \psi_\Lambda) = \lambda (\omega^P \psi_\Lambda)$$

and, therefore, an eigenvalue of $A^P$ itself. We conclude that the set $\{\lambda_P\}$ of eigenvalues of $A^P$ is a subset of the set $\{\lambda_\Lambda\}$ of eigenvalues of $\Lambda$. The process can be repeated with a subgroup Q of P. The linear operator $A^Q$ associated with Q is given by

$$A^Q = \frac{1}{n_Q} \sum_Q \omega^Q A^P \omega^{Q-1}$$

The set $\{\lambda_Q\}$ of the eigenvalues of $A^Q$ is a subset of $\{\lambda_P\}$. We can consider all possible consecutive subgroups until there is no more subgroup left. We thus obtain a hierarchy of eigenvalues that completely characterize the eigenstates of the system.

Let $\psi_P$ and $\lambda_\rho$ be a complete set of eigenvectors and eigenvalues of $A^P$.

$$A^P \psi_P = \lambda_P \psi_P$$

One has

$$A^P = \sum_P \psi_P \lambda_P \psi_P^{T*}$$

and the closure relation

$$1 = \sum_\rho \psi_P \psi_P^{T*}$$

The physical states $\psi$ are normalized

$$\psi^{T*} \psi = \sum_P (\psi^{T*} \psi_P)(\psi_P^{T*} \psi) = \sum_P \varpi_P = 1$$

The $\varpi_P$'s are positive numbers and their sum is 1. They may be considered therefore as probabilities. The average value of $A^P$ in state $\psi$, computed by using these probabilities, is given by

$$\langle A^P \rangle = \sum_P \varpi_P \lambda_P / \sum_P \varpi_P = \sum_P \psi^{T*}\left(\psi_P A^P \psi_P^{T*}\right)\psi / \left(\psi^{T*}\psi\right) = \frac{\psi^{T*} A^P \psi}{\left(\psi^{T*}\psi\right)}$$

We so obtain the second postulate of quantum theory:
To each dynamical variable (a physical concept) there corresponds a linear operator $A$ (a mathematical object), or observable, that operates in the Hilbert space. The average value of an operator $A$ for a system in state $\psi$ is given by

$$\langle A \rangle = \frac{\psi^T * A \psi}{\psi^T * \psi}$$

C. Third postulate

The third postulate determines the dynamics of a quantum system: The dynamics of a quantum system is given by a Schrödinger equation

$$i\hbar \frac{d\psi(t)}{dt} = H\psi(t)$$

where $H$ is an operator called the Hamiltonian of the system.
The quantum state $\psi_i$ at a world point 'i' is expressed in terms of its polarization components:

$$\psi_{i,\mu} = \sum_\nu C_{\mu\nu} \varphi_{i\nu}$$

where $C$ is a 4-dimensional matrix that operates in the internal space of world points
Let us now look more carefully at the expression of the Lagrange operator

$$\Lambda = \Delta \otimes G$$

Any square matrix such as $\Delta$ can be expressed, according to the LDU theorem of Banachiewicz, as a product of a lower triangular matrix $L$, a diagonal matrix $A$ and an upper triangular matrix $U$. When the matrix is real and symmetric, as is the case for $\Delta$, the two triangular matrices are each other transpose:

$$\Delta = D^T A D$$

where $D$ is a triangular, here a random triangular matrix, that is $D_{ij}=0$ for $i<j$. $D^T$ is the transpose matrix and $A$ is diagonal. More precisely, since space is homogeneous, $A$ is a spherical matrix (it is proportional to the unity matrix: $A_{ij}=a\delta_{ij}$) and one may take $a=1$. Hence

$$\Delta = D^T D$$

The elements $D_{ij}$ of $D$ are sums of $n$ binary random variables $\pm J/n$. Their distribution is Gaussian and given by

$$P(D_{ij}) = \frac{1}{J\sqrt{2\pi/n}} \exp\left(-\frac{D_{ij}^2}{2J^2/n}\right)$$

The off-diagonal elements are of the order of $D_{ij} \cong J/\sqrt{n}$. $D$ is made dimensionless by dividing all its elements by $-J$. The interaction strength $J$ is transferred to $G$. Then $D_{ij} \cong \pm 1/\sqrt{n}$. The elements $\Delta_{ik}$ of $\Delta$ are given, for $i \neq k$, by

$$\sum_j D_{ij}^T D_{jk} = \Delta_{ik}$$

and for $i = k$ by

$$\sum_j D_{ij}^T D_{ji} = \sum_j D_{ij}^2 = \Delta_{ii}$$

Since $D_{ij}^2 \cong 1/n$ one has $\Delta_{ii} \cong 1$.

One defines the increment of a polarization component $\varphi_{i\mu}$ of world point "$i$" by

$$\delta\varphi_{i\mu} = \sum_j D_{ij}\varphi_{j\mu}$$

that is

$$\delta\phi_i = \sum_j D_{ij}\phi_j$$

The increment of the $v^{th}$ component of the vector field $\psi_i$ along dimension $\mu$ is written as

$$\delta_\mu \psi_{iv} = C_{\mu v}\delta\varphi_{iv} = C_{\mu v}\sum_j D_{ij}\varphi_{jv}.$$

In appendix A2 we show that he operator $D$ may be seen as a differential operator indeed. The partial derivatives of the quantum field components are given by

$$\partial_\mu \psi_{iv} = \frac{1}{l*}\delta_\mu \psi_{iv} = \frac{C_{\mu v}}{l*}\sum_j D_{ij}\varphi_{jv}$$

for the first order derivatives, and by

$$\partial_\mu^2 \psi_{iv} = \frac{C_{\mu v}}{l*^2}\sum_{jk} D_{ij}^T D_{jk}\varphi_{kv}$$

for second order derivatives. It is reminded that $l*$ is the metric limit that is the size of a world point. By introducing the coefficient $C_{\mu v}$ into one entry the eigenvalue equation

$$(\Delta \otimes G)\psi = \kappa\psi$$

one finds

$$C_{\mu v}\sum_{jk,v} D_{ij}^T G_{0\mu v} D_{jk}\varphi_{kv} = \kappa C_{\mu v}\varphi_{i\mu}.$$

With the definition of second order derivatives this expression writes

$$\varepsilon_\mu G_{0\mu} l*^2 \partial_\mu^2 \psi_{iv} = \kappa C_{\mu v}\varphi_{i\mu}.$$

We recall from the previous section that $G_0$ has one time-like and three equivalent space-like dimensions. The summation over index $\mu$ is then carried out on both members of the expression. By introducing the Minkowski metrics ($\varepsilon_t = -1$ ; $\varepsilon_r = 1$) one finds

$$\left[-G_{0t}l*^2 \frac{\partial^2 \psi_v}{\partial t^2}\right] + G_{0r}l*^2 \Delta\psi_v = \kappa\psi_v \qquad (7)$$

where $\Delta$ is here the usual three dimensional Laplacian. The parameters c and $\hbar$ that have been defined earlier, are introduced in that expression; c, called the velocity of light, is given by

$$c^2 = \frac{G_{0r}}{G_{0t}} = \frac{1}{bJ-1}$$

It is, in fact, a dimensionless, universal, parameter that determines the ratio between the standards of length and time. $\hbar = l*G_{0r}^{1/2}$ is called the Planck constant. Finally one defines the mass of the particle associated with the field $\psi$ by $\kappa = (mc)^2$ (provided that $\kappa > 0$). The equation (7) becomes

$$\left(\frac{1}{c^2}\frac{\partial^2}{\partial t^2} - \Delta + \left(\frac{mc}{\hbar}\right)^2\right)\psi_v(r,t) = 0$$

which is recognized a set of four Klein-Gordon equations. The introduction of a potential $V$ modifies the equation

$$\left(\frac{1}{c^2}\frac{\partial^2}{\partial t^2} - \Delta - V + \left(\frac{mc}{\hbar}\right)^2\right)\psi_\nu(r,t) = 0$$

In the non-relativistic limit, where the mass term $mc^2$ is much larger than the other terms, the equation may be approximated by

$$(i\hbar)^2 \frac{\partial^2 \psi}{\partial t^2} \cong \left[mc^2 - \hbar^2\left(\frac{\Delta}{2m} - \frac{V}{2m}\right)\right]^2 \psi$$

whose solution is

$$i\hbar \frac{\partial \psi}{\partial t} \cong \pm\left(mc^2 - \hbar^2\left(\frac{\Delta}{2m} - \frac{V}{2m}\right)\right)\psi = \pm H\psi$$

with

$$H = mc^2 - \hbar^2\left(\frac{\Delta}{2m} - \frac{V}{2m}\right)$$

The + sign corresponds to Schrödinger equation and we recover the third postulate.
In conclusion the set of postulates of quantum theory may be seen as a consequence of the model of discrete space that we put forward in this essay.

## 4. THE ORGANIZATION OF ELEMENTARY PARTICLES

### A. Bosons and fermions

A quantum system, comprised of $N$ world points, contains a particle $P$ if the matrices $G$ associated with the internal spaces of the world points of the system are all identical to $G_0$ the vacuum metrics except for one world point $a$ where $G = G_P$, a matrix characterized by the symmetry properties of $P$. The possible physical states of the quantum system are then solutions of the following eigenvalue equation

$$\left(\Delta \otimes G^{(P)}\right)\psi_\kappa = \kappa_P \psi_\kappa \qquad (8)$$

where $G^{(P)} = \{G_i\}$ is the set of all matrices of the quantum system. The symmetry of fundamental particles is traditionally associated with the irreducible representations of group $S_2$, the group of permutations of two particles. This group has two elements and two irreducible representations namely the even representation that is associated with the permutation properties of pairs of bosons and the odd representation that is associated with the permutation properties of pairs of fermions. This presentation is not satisfactory, however, for two reasons

First, in this definition, the bosonic or fermionic character seems to be a property of pairs of particles and an isolated fermion, for example, is not defined.

Secondly, it does not fit the idea that we support, that the nature of particles relates to their individual symmetry properties. In fact this problem has already been pointed out by theoreticians who believed that the symmetry of fundamental particles ought to be searched in the irreducible representations of a fundamental group and put forward the Poincaré's (or Lorentz) group. Our idea is close to this proposal with the difference that the relevant group would be the local group $S_4$ of permutations of 4 objects instead. The table of characters of $S_4$ has been given in Table I.

The 5 irreducible representations of group $S_4$ write

$$\Gamma_1, \Gamma_1^*, \Gamma_2, \Gamma_3, \Gamma_3^*$$

with orders 1, 1, 2, 3 and 3 respectively. The irreducible representations $\Gamma_1$ and $\Gamma_3$ are insensitive to mirror operations while $\Gamma_1^*$ and $\Gamma_3^*$ are sensitive to mirror transformations (triggered by an odd number of permutations). One has $\Gamma_3^* = \Gamma_1^* \Gamma_3$ and $\Gamma_2^* = \Gamma_1^* \Gamma_2 = \Gamma_2$. $G_P$ must commute with any four dimensional representation of group $S_4$ and therefore it must transform according to direct sums of irreducible representations of $S_4$.

Let us first try to build the matrix $G_P$ by using $\Gamma_1$, $\Gamma_2$ and $\Gamma_3$.

The first possibility is to build $G_P$ as a direct sum of $\Gamma_1$ and $\Gamma_3$

$$G_B \approx \Gamma_1 \oplus \Gamma_3 \qquad (9)$$

This corresponds to the solution 4=1+3. Cosmic points that are polarized along that symmetry display a *bosonic polarization*. Accordingly a bosonic polarization is described by a quadrivector.

The other possibility is to build $G_P$ on the representation $\Gamma_2^* (\equiv \Gamma_2)$.

$$G_F \approx \Gamma_2 \oplus \Gamma_2 = 1^{(2)} \otimes \Gamma_2 \qquad (10)$$

This corresponds to the solution 4=2+2. Cosmic points that are polarized along that symmetry display a *fermionic polarization*. The representation $\Gamma_2$ determines the fermionic, or spinor, character of the polarization. Since a fermionic polarization implies this representation $\Gamma_2$ twice, it is called a bispinor.

### B. A family of elementary particles

Super-symmetry theory (SUSY) puts forward that fermions and bosons may be considered as two aspects of the very same objects. In a spirit close to that of SUSY it is suggested here that the fundamental particles are objects that associate world points with different symmetries. A particle would be made of a pair of coupled world points, one undergoing a fermionic polarization (10) and the other a bosonic polarization (9).

According to this model the state of a particle is represented in a 16-dimensional vector space that is obtained by the direct product of the two 4-dimensional spaces associated with the members of the pair.

The states must therefore transform as

$$(\Gamma_1 \oplus \Gamma_3) \otimes (\Gamma_2 \oplus \Gamma_2)$$

which may be expanded along

$$[\Gamma_1 \otimes \Gamma_2] \oplus [\Gamma_1 \otimes \Gamma_2] \oplus [\Gamma_3 \otimes \Gamma_2] \oplus [\Gamma_3 \otimes \Gamma_2]$$

Every bracket is related to a given type of particle *P*. As a result a family of particles is comprised of four types of particles. This fits the Standard Model. Moreover all these particles have a fermionic character since they all transform according to $\Gamma_2$. One also observes that there are two classes of particles.

-a)- The first class is associated with the two representations that mix $\Gamma_1$ with $\Gamma_2$. It may characterize the lepton particles. The particle associated with one of the two representations is a charged lepton, that associated with the other representation is the associated neutrino.

-b)- The other class associates $\Gamma_3$ with $\Gamma_2$: It may characterize the quark particles. The particles associated with one of the two representations are then the three colour versions of one quark, the particles associated with the other representation are the three colour versions of the other quark.

The physical size of a particle would be $2l^*$ where $l^*$ is the size ($l^* \cong 0.5 \times 10^{-21}$ cm) of a world point. This is the smallest segment that may be given a physical meaning. It is possibly related to the minimal segment that arises from the non commutative geometry theory of the Standard Model of Connes.

If, instead of using the mirror insensitive representations $\Gamma_1$ and $\Gamma_3$ one appeals to the mirror sensitive representations $\Gamma_1^*$ and $\Gamma_3^*$ of $S_4$ one obtains another set of four particles, actually four antiparticles. The states of anti-particles transform as

$$(\Gamma_1^* \oplus \Gamma_3^*) \otimes (\Gamma_2 \oplus \Gamma_2)$$

that may be expanded along

$$[\Gamma_1^* \otimes \Gamma_2] \oplus [\Gamma_1^* \otimes \Gamma_2] \oplus [\Gamma_3^* \otimes \Gamma_2] \oplus [\Gamma_3^* \otimes \Gamma_2]$$

If $\psi_P$ is the quantum state of a system that contains a particle, $\overline{\psi_P}$ the quantum state of the associated antiparticle is simply given by $\overline{\psi_P} = -\psi_P = C\psi_P$. C is the charge conjugation operator. The four antiparticles have the same physical properties as the particles except for their electric charges. For a discussion on the CPT theorem see Appendix A3.

We have so far obtained a convenient description of the organization of one family of elementary particles but one knows that there exist three families of particles with identical properties except for the masses. Since the symmetry properties are the same for the three families one must admit that the matrices $G_P$ are the same for the three families. The particles of the first family, however, are stable whereas the particles belonging to the other families are unstable. Therefore the existence of these families cannot be looked for in some new type of symmetry similar to SU(3) for example. At this point the question remains open.

## 5. DISCUSSIONS AND CONCLUSIONS

In this first contribution we put forward a model of space-time where we assume that the universe as a whole is made of elementary physical objets, called cosmic bits, whose states are entirely determined by one, and only one, binary variable. The idea that the universe is made of bits is not new. Wheeler, for example, states that physics at large could be understood in terms of '*It from bit*'(2). There is however a fundamental difference between his approach and ours. In the Wheeler approach the bits are to be understood as bits of information that is signals transmitted from an emitter to a receiver. The physical laws are the results of computations carried out on those bits by a huge sort of universal computer, a Turing machine for example, according to convenient programs. The physical world would be the result of these computations and the physicists would be the receivers. In our approach there is no program and no programmer behind the stage. The bits are physical objects, not signals, that constitutes a system somehow similar to a ferromagnetic powder. The process that moves the bits is purely physical and determined by statistical physics. Moreover in our approach time and space are treated on equal footing, in the spirit of relativity theory, so avoiding the philosophical problems arising from the necessary existence of a clock driving the computer.

Obviously the difficulty is to show that the general laws of physics can be recovered from so simple a model. It seems that the difficulty can be overcome in most cases. In this contribution the structure of space-time has been recovered, the main postulates of quantum theory have been reestablished and the organization of elementary particles shown to fit that of the Standard Model of particles. More remains to be done, for example to describe the Standard Model with more details or to show how gravitation and general relativity come into play. Those topics will be discussed in other contributions.

APPENDIX A1: The partition function of world points W.

The Lagrangian of a world point W made of $d$ sub-points is given by
$$\Lambda(W) = -\frac{J}{2n}\sum_{\alpha\mu,\beta\nu}\sigma_{\alpha\mu}\sigma_{\beta\nu}$$

With
$$s_\mu = \frac{1}{n_\mu}\sum_\alpha \sigma_{\mu\alpha}$$

one has
$$n_\mu n_\nu s_\mu s_\nu = \left(\sum_\alpha \sigma_{\mu\alpha}\right)\left(\sum_\beta \sigma_{\nu\beta}\right) = \sum_{\alpha\beta}\sigma_{\mu\alpha}\sigma_{\nu\beta}$$

and
$$\Lambda = -\frac{J}{2n}\sum_{\mu\nu} n_\mu n_\nu s_\mu s_\nu$$

The partition function is
$$Z = \sum_{\{\sigma\}} \exp(-b\Lambda)$$

where the sum is over all possible configurations of the world point. The sum may be carried out in two steps
$$Z = \sum_{\{s_\mu\}} \exp\left(\frac{bJ}{2n}\sum_{\mu\nu} n_\mu n_\nu s_\mu s_\nu\right) \sum_{\{\sigma\}\,s_\mu\,\text{given}} 1$$

that is a sum over all possible sets of polarizations and, then, a sum over all bits states for a given set of polarizations. The last term
$$w = \sum_{\{\sigma\},s_\mu\,\text{given}} 1$$

is purely combinatorial in nature. It is given by
$$w = \frac{n!}{\prod_\mu (n_{\mu\uparrow}!\, n_{\mu\downarrow}!)}$$

where
$$n_{\mu\uparrow} = \frac{n_\mu}{2}(1+s_\mu)$$
$$n_{\mu\downarrow} = \frac{n_\mu}{2}(1-s_\mu).$$

By using the Stirling formula one obtains
$$\text{Ln}(w) \cong -\sum_\mu \left[\frac{n_\mu}{2}\left((1+s_\mu)\text{Ln}(1+s_\mu) + (1-s_\mu)\text{Ln}(1-s_\mu)\right)\right]$$

where a non relevant constant has been skipped. The partition function reads:
$$Z = \sum_{\{s_\mu\}} \exp\left[\frac{Jb}{2n}\sum_{\mu\nu} n_\mu n_\nu s_\mu s_\nu - \sum_\mu \left[\frac{n_\mu}{2}\left((1+s_\mu)\text{Ln}(1+s_\mu) + (1-s_\mu)\text{Ln}(1-s_\mu)\right)\right]\right]$$

a sum that, in the thermodynamic limit, reduces to one term where $s_\mu \equiv \langle s_\mu \rangle = \varphi_\mu$ ( the so-called saddle point method). $F$, the free energy, is defined by. $Z = \exp(-bF(\varphi_\mu))$ One has with $n_\mu \cong n/d$

$$F = \frac{-nJ}{2d^2}\sum_{\mu\nu}\varphi_\mu\varphi_\nu + \frac{n}{2bd}\sum_\mu\left((1+\varphi_\mu)\mathrm{Ln}(1+\varphi_\mu)+(1-\varphi_\mu)\mathrm{Ln}(1-\varphi_\mu)\right)$$

The realizable physical states are those that minimize $F$

APPENDIX A2: The Leibniz formula for operator $D$.

For $D$ to be a differential operator it is necessary to show that it is linear and that it obeys the Leibniz formula.
Let us consider two quantum states $\phi$ and $\eta$. One has

$$\sum_j D_{ij}(\phi_j + \eta_j) = \sum_j D_{ij}\phi_j + \sum_j D_{ij}\eta_j$$

that is $\delta(\phi+\eta) = \delta\phi + \delta\eta$ and $D$ is linear indeed.
On the other hand

$$\sum_j D_{ij}\phi_j\eta_j = \sum_j D_{ij}\left[(\phi_j - \phi_i)+\phi_i\right]\left[(\eta_j - \eta_i)+\eta_i\right]$$
$$= \sum_j D_{ij}(\phi_i\eta_j + \phi_j\eta_i) - \sum_j D_{ij}\phi_i\eta_i + \sum_j D_{ij}\left[(\phi_j - \phi_i)(\eta_j - \eta_i)\right]$$

The second term vanishes because the elements of $D$ are random

$$\sum_j D_{ij}\phi_i\eta_i = \phi_i\eta_i\sum_j D_{ij} \cong 0$$

The third term is a second order term. Both may be ignored and one has

$$\sum_j D_{ij}(\phi_i\eta_j + \phi_j\eta_i) = \phi_i\sum_j D_{ij}\eta_j + \eta_i\sum_j D_{ij}\phi_j$$

That is $\delta(\phi\eta) = \phi\delta\eta + \eta\delta\phi$, the Leibniz formula.

APPENDIX A3: On the CPT theorem

The vector

$$\phi = \begin{pmatrix}\varphi_1 \\ \varphi_2 \\ \varphi_3 \\ \varphi_4\end{pmatrix}$$

represents the state of a particle. The state $\bar{\phi}$ of the associated anti-particle is obtained from $\phi$ through a mirror transformation that is a sign change of one of its components. Nothing, however, determines which component has to be modified and, therefore, all signs have to be changed. We note that if the polarization $\varphi$ of a world point is a solution of the equation $\varphi = \tanh(bJ\varphi)$ then $-\varphi$ is also a solution of the equation. Whence

$$\bar{\phi} = \begin{pmatrix} -\varphi_1 \\ -\varphi_2 \\ -\varphi_3 \\ -\varphi_4 \end{pmatrix} = \begin{pmatrix} -1 & & & \\ & -1 & & \\ & & -1 & \\ & & & -1 \end{pmatrix} \begin{pmatrix} \varphi_1 \\ \varphi_2 \\ \varphi_3 \\ \varphi_4 \end{pmatrix} = C\phi$$

C is the charge conjugation operator.

In other respects the time reversal operator T changes the sign of the time component $\varphi_1$ of $\phi$ that is

$$T \begin{pmatrix} \varphi_1 \\ \varphi_2 \\ \varphi_3 \\ \varphi_4 \end{pmatrix} = \begin{pmatrix} -\varphi_1 \\ \varphi_2 \\ \varphi_3 \\ \varphi_4 \end{pmatrix} \qquad T = \begin{pmatrix} -1 & & & \\ & 1 & & \\ & & 1 & \\ & & & 1 \end{pmatrix}$$

Finally the operation of reversing the direction of one axis, say the z direction, amounts to a reflection of space (a mirror symmetry operation) in any number of space dimensions, and in three space dimensions it is equivalent to reflecting all space coordinates, because one can add an additional rotation of 180 degrees in the x-y plane to complete the transformation. Let P be the operator that reflects all space coordinates. P amounts to changing the signs of the three spatial components $\varphi_2$, $\varphi_3$ and $\varphi_4$.

$$P \begin{pmatrix} \varphi_1 \\ \varphi_2 \\ \varphi_3 \\ \varphi_4 \end{pmatrix} = \begin{pmatrix} \varphi_1 \\ -\varphi_2 \\ -\varphi_3 \\ -\varphi_4 \end{pmatrix}$$

that is

$$P = \begin{pmatrix} 1 & & & \\ & -1 & & \\ & & -1 & \\ & & & -1 \end{pmatrix}$$

We see that

$$CPT = 1^{(4)} = \begin{pmatrix} 1 & & & \\ & 1 & & \\ & & 1 & \\ & & & 1 \end{pmatrix}$$

a simple and straightforward derivation of the CPT theorem.

References
(1) P. Peretto, Eur. Phys. J. **C 35**; 567 (2004)
(2) J. Wheeler in '*Complexity, Entropy and the Physics of Information*' Zurek ed. (Addison-Wesley 1996)